# A mixed model approach to drought prediction using artificial neural networks: Case of an operational drought monitoring environment

Chrisgone Adede*, Robert Oboko, Peter Wagacha and Clement Atzberger**

*School of Computing and Informatics, University of Nairobi (UoN), P.O. Box 30197, GPO, Nairobi, Kenya

**Institute of Surveying, Remote Sensing and Land Information, University of Natural Resources (BOKU), Peter Jordan Strasse 82, A-1190 Vienna, Austria

* Correspondence: Tel.: +254-724-550-746; E-Mail: adedekris@gmail.com

**Abstract:**

Droughts, with their increasing frequency of occurrence, continue to negatively affect livelihoods and elements at risk. For example, the 2011 in drought in east Africa has caused massive losses document to have cost the Kenyan economy over $12bn. With the foregoing, the demand for ex-ante drought monitoring systems is ever-increasing. The study uses 10 precipitation and vegetation variables that are lagged over 1, 2 and 3-month time-steps to predict drought situations. In the model space search for the most predictive artificial neural network (ANN) model, as opposed to the traditional greedy search for the most predictive variables, we use the General Additive Model (GAM) approach. Together with a set of assumptions, we thereby reduce the cardinality of the space of models. Even though we build a total of 102 GAM models, only 21 have $R^2$ greater than 0.7 and are thus subjected to the ANN process. The ANN process itself uses the brute-force approach that automatically partitions the training data into 10 sub-samples, builds the ANN models in these samples and evaluates their performance using multiple metrics. The results show the superiority of 1-month lag of the variables as compared to longer time lags of 2 and 3 months. The champion ANN model recorded an $R^2$ of 0.78 in model testing using the out-of-sample data. This illustrates its ability to be a good predictor of drought situations 1-month ahead. Investigated as a classifier, the champion has a modest accuracy of 66% and a multi-class area under the ROC curve (AUROC) of 89.99%

## 1. Introduction

A drought is a recurrent event marked by lack of precipitation for extended period of times (Morid, Smakhtin & Bagherzadeh, 2007; Bordi et al, 2005) [1,2]. Droughts are one of the most complex and less understood disasters, having the greatest impacts on people and usually affecting large regions (Morid, Smakhtin & Bagherzadeh, 2007; Ali et al.,2017) [1,3].

Droughts have the types indicated in Figure 1. Common to all types is their slow onset and that we have to deal with a gradually progressing phenomena. The progression is characterized by deficiency of precipitation, effects on surface to sub-surface water sources, followed by reduced vegetation growth and finally a culmination on socio-economic effects on people and livelihoods UNOOSA (2015) [4].

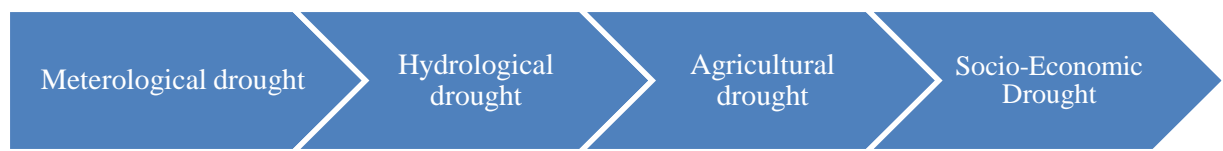

**Figure 1:** Conceptualization of various drought types and their progression. The impacts on people and livelihoods is a function of the vulnerability of the livelihoods as well as the severity, duration and spatial extent of the drought in relation to the local at-risk elements within the drought hazard.

There is an increase in both the frequency of drought and the cost of economic losses as a result of droughts throughout the world and particularly in Kenya and East Africa. For example, Government of Kenya (2012) [6] documents the 2008-2011 drought in Kenya as having made 3.7Million people food insecure with economic losses approximated at US$12.1 billion. The 2014 California drought was projected to have cost a total of $2.2 billion in losses Cody (2010) [7]. Further documentation of losses, especially of household due to drought are reviewed in Udmale et al (2015) [6] and in Ding, Hayes & Widhalm (2011) [8].

The spike in drought related losses has led to the focus on Drought Risk Management (DRM) systems. The key elements of a DRM system are drought risk identification, drought monitoring, drought preparedness, and drought mitigation. While risk identification involves the agreement on the definition of droughts based on some objective parameters, drought risk monitoring is based on the establishment of appropriate drought early warning systems (EWS) that signal the advent, progression and even possible cessation of drought events. As advocated by Mariotti et al (2013) [9], drought risk identification and drought early warning systems are the starting points to sound drought risk management that can greatly reduce severity of social and economic damage by droughts. Drought risk mitigation aims to reduce on impacts of on-course droughts and, in the Kenyan context, is through two main approaches. First is the disbursement of social security funds to the population in the four most venerable counties of Turkana, Marsabit, Mandera and Wajir as outlined in Beesley (2011) [10]. These counties comprise our study area. Second is the Drought Contingency Fund (DCF) that are used to provide essential services across the sectors of livestock, water, health and nutrition, education and security.

Drought early warning systems (EWS) are in most cases based on remote sensing data and in some cases on incorporation of socio-economic data to measure the impacts of droughts (Rembold et al., 2013) [31]. The remote sensing data in those EWS is essential as it permits a cost-effective spatio-temporal mapping of vegetation conditions and crop yield (Atzberger, 2013). Indices are mostly derived from either of precipitation data or observed vegetation conditions. Some of the indices includes rainfall anomaly-based indices like Rainfall Condition Index (RCI), Crop Moisture Index (CMI), Standardized Precipitation Index (SPI), Standardized Precipitation-Evapotranspiration Index (SPEI) and vegetation-based indices like Normalized Difference Vegetation Index (NDVI), Vegetation Condition Index (VCI), and Standardized Vegetation Index (SVI). The use of SPI in drought monitoring and also in drought forecasting is well documented in Bordi et al (2005) [2], Huang et al (2016) [12], Khadr (2016) [13] and Wichitarapongsakun et al (2016) [14]. Klisch & Atzberger (2016) [18] document the drought monitoring system for Kenya as implemented by the University of Natural Resources and Life Sciences, Vienna (BOKU). The system is premised on both indices based on both vegetation and precipitation. The indices include NDVI and its derivatives of VCI aggregated over 1- and 3-months periods and SPI aggregated over the same periods.

Most of the drought early warning systems (EWS), are either near real time (NRT) [15-18] or ex-post and thus provide information at the lapse of the periods of monitoring. NRT approaches are documented in Hayes et al (1999) [15], AghaKouchak (2012) [17] and recently Klisch & Atzberger (2016) [18]. The greatest limitation to having EWS on a near real time basis is processing latencies that come with both satellite-based and ground sourced data. For example, the widely used FEWSNET drought indicators are delivered only with a lag time of one month, necessary to complete the data pre-processing. Near real time (NRT) approaches like in Klisch & Atzberger (2016) [18] that provides vegetation indices without latency are currently most desired.

The above traditional approach to drought management in which droughts are monitored and responded to as and after they unravel, and the learning on the losses from past drought events, has led to a shift towards Drought Risk Management (DRM). DRM is considered a holistic approach that includes the formulation of drought sensitive policies at sub-national, national and even regional levels. Key elements are drought contingency planning, drought early warning, drought resilience building and preparedness, drought impact assessment, drought communication, drought response and drought recovery. One key tenet of the DRM approach is the need for reliable prediction driven systems and models that are incorporated as part of the drought EWS. Drought prediction in its simplest form involves any attempts to forecast an outlook of how the future of drought situations will be. These pointers then form the basis for early prepared measures towards mitigation, response and resilience building. The increasing need for the incorporation of prediction capabilities to DRM models is for example documented by Mariotti et al (2013) [9] and recognises that the task of forecasting is currently below user needs implying lag on the supply side especially at regional, national and local scales.

Drought prediction approaches to DRM have been variously studied. The defining differences can be viewed through the twin lenses of the data used and the methods deployed. Most of the approaches are based on the use of single variables of either precipitation or vegetation conditions. Examples of predictions based on precipitation data are either based on SPI as is the case in Ali et al. (2016)[3], Huang et al. (2016)[12] and Khadr (2016) [13], or on river flow indices like Yuan et al. (2017) [22]. While some of the studies are based on SPEI as either the main indicator or in addition to SPI as documented in Morid, Smakhtin, & Bagherzadeh(2007) [1], Le et al.(2016) [], Maca and Pech [2-16], [1,2,20,24] others define a super index of drought indices in the approach of [23] and [25] that define Multi-variate standardised dry index (MSDI) and Drought defining Index (DDI) respectively. The use of vegetation conditions in [21] in a forecast study stands-out in its use of 11 attributes to predict vegetation conditions. This is as opposed to say [18] that directly predict vegetation conditions based on measured data.

In terms of methods and approaches, most of the studies focus on a single method. Employed methods encompass purely statistical approaches to Machine Learning (ML) techniques. Statistical methods have a range in complexity from simple forecasting to Multiple Regression Techniques and to ensemble approaches that employ more than one modelling technique. The ML methods range from Neural Networks in Morid, Smakhtin & Bagherzadeh (2007) [1], Ali et al. (2017) [3] Le et al. (2016) to Regression Tree Techniques (RTT) in Tadesse et al. (2016) [21], Hidden Markov Models in Khadr (2016) [13] and Kalman filters (Sedano, Kempeneers and Hurtt, 2014) [26].

In this study, we present multi-variate drought predictive model that uses combination of two techniques of two methods, statistical and machine learning, to predict drought conditions up to 3 months ahead. The approach also evaluates and selects champion model from the space of all possible models based on objective evaluation metrics.

## 2. Material and Methods

*2.1 Study Area*

The study area is shown in Figure 1. The study area comprises four counties of Kenya: Turkana, Marsabit, Mandera and Wajir. The selected region lies in the northern part of Kenya that is characterized as part of the arid and semi-arid lands (ASALs) of Kenya. The selected counties are classified as arid and part of the ASALs monitored by the National Drought Management Authority (NDMA) of Kenya.

The four counties cover a combined area of 215,242 km$^2$ with a total population of around 2.8M. The annual average rainfall is 250mm (Turkana, Marsabit and Mandera) to 370mm (Wajir). The monthly average vegetation cover (2003-2015), as quantified by the Normalized Difference Vegetation Index (NDVI) from NDMAs operational drought monitoring system, is shown in Figure 2. The low values indicate sparse vegetation cover even during the wettest months.

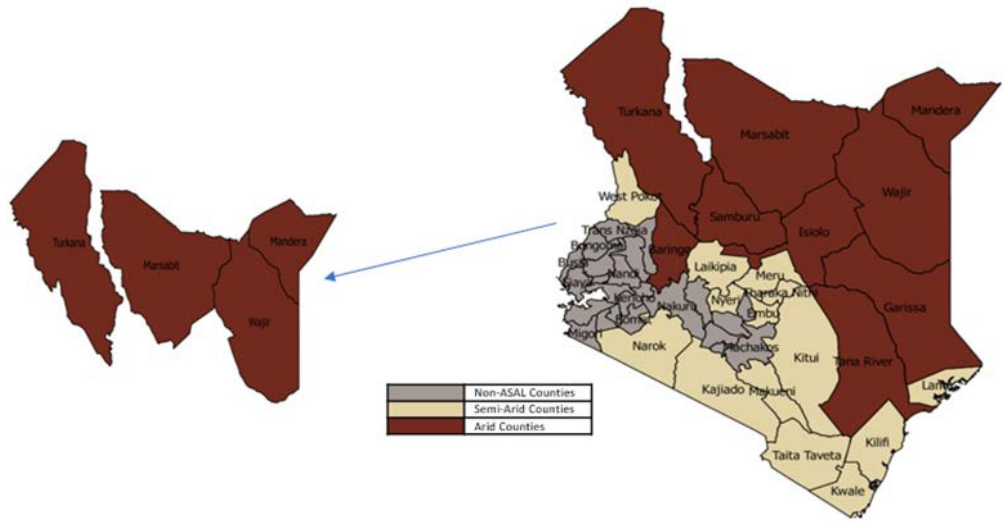

**Figure 1:** Study area of counties Turkana, Marsabit, Mandera and Wajir in Kenya. The map of Kenya to the right is showing the grouping of its counties into arid, semi-arid and non-ASALs. The extent is bounded by UL X (Lon) 33.918, UL Y (Lat) 5.513, LR X (Lon) 41.967 and LR Y (Lat) 0.147.

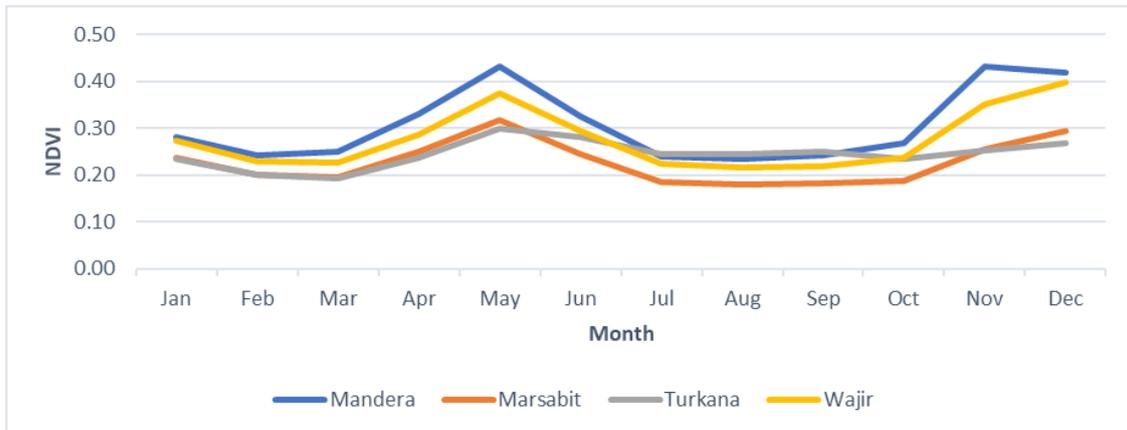

**Figure 2:** Average NDVI (2003-2015) across months by county based on NDMAs operational EWS.

The classification of drought based on 3-monthly aggregated Vegetation Condition Index (VCI) following on the thresholds used in Klisch & Atzberger (2016) [18] over the period March 2001-December 2015 (178 months) shows why the problem of droughts is quite immense for the study area (Table 1).

**Table 1:** Summary of monthly drought phases for the counties in the study region (03-2001 to 12-2015). The 3-monthly aggregated VCI is classified according to thresholds proposed in Klisch & Atzberger (2016). The 3-monthly VCI is updated every 10 days leading to 712 possible episodes.

| County | Extreme | Severe | Moderate | Combined |
|---|---|---|---|---|
| Mandera | 8 | 31 | 43 | **82** |
| Marsabit | 8 | 26 | 70 | **104** |
| Turkana | 4 | 28 | 64 | **96** |
| Wajir | 9 | 25 | 61 | **95** |
| **Total** | **29** | **110** | **238** | **377** |

Over the 178 months period, 377 out of a possible 712 (52%) drought episodes are reported at county level, 29 (4%) of which being classified as extreme (VCI<10) and therefore signalling possible collapse of community coping capabilities. Early warning systems with predictive/ forecasting capabilities is thus a possible value addition to the drought monitoring process for the counties in this study area. The poverty levels in these counties make drought effects even more impactful on the communities due to high vulnerabilities.

*2.2 Data*

Despite the differences in the formulation of the predictive modelling steps, the common broad steps can be grouped into three stages: pre-modelling, model building and model deployment stages. The pre-modelling stage involves all the steps at definition of the modelling objective, data acquisition, data preparation, variable selection and variable transformation. The model building stage involves the formulation of multiple models, their evaluation, validation and subsequent selection. In most cases, the above processes and stages are not linear but iterative in the search for the optimum model as provided by the best set of predictors. We discuss the pre-modelling stage in this section while the model building and deployment stages are discussed in section 2.3

The variables used in this predictive study comprises both precipitation and vegetation indices with either one month or three-month aggregation periods. The precipitation datasets are derived from TAMSAT and includes Rainfall Estimates (RFE), Rainfall Condition Index (RCI) and Standardized Precipitation Index (SPI). Vegetation conditions are characterized through Normalized Difference Vegetation Index (NDVI) and Vegetation Condition Index (VCI) directly provided by BOKU. The choice to use the BOKU dataset is based on the fact that the same is deployed in Kenya's operational drought monitoring system at NDMA. The data is sourced from the NDMA monitoring systems as deployed by BOKU and administered by the author. Several studies evaluated the BOKU dataset including comparisons against similar products (Atzberger et al, 2016[30]).

Klisch & Atzberger (2016) [18] and others document the use of Vegetation Condition Index (VCI) as a temporal and spatially aggregated anomaly of the Normalised Difference Vegetation Index (NDVI). While the NDVI gives absolute vegetation status for a given spatial extent at a given time, the VCI scales the actual NDVI value in the range between a historical minimum (VCI = 0%) and maximum (VCI = 100%) for a given time unit. Widely used time units are dekads (10 day periods), months and 3 months periods.

The above precipitation and vegetation related indices are calculated at pixel level then aggregated over the appropriate time-scales and administrative boundaries. The details and formulae for the computation of these indices are as provided in Table 2.

**Table 2:** A description of the index calculation formulas. $NDVI_i$ indicates the NDVI observed at time i; $NDVI_{min}$ and $NDVI_{max}$ are minimum and maximum NDVI observed in the period 2003-2013. nIR and Red are the spectral reflectances in near infrared and red spectral channels of MODIS satellite, respectively. Before use, the NDVI time series is smoothed and filtered to remove negative impacts of poor atmospheric conditions and undetected clouds [18]

| Variable /Index | Index Calculation | Index Description |
|---|---|---|
| VCI | $VCI_{c,i}=100* (NDVI_{c,i}-NDVI_{min\ c,i})/ (NDVI_{max\ c,i}- NDVI_{min\ c,i})$ | Predicted variable aggerated over 3 months period. |
| NDVI | $NDVI = (nIR – Red) ./ (nIR +Red)$ | Predictor variable; measures the average monthly vegetation greenness. |
| RFE | Rainfall estimate from TAMSAT product (in mm) | Predictor variable; estimate of the monthly rainfall |
| RCI | $RCI_{c,i}=100* (RFE_{c,i}-RFE_{min\ c,i})/ (RFE_{max\ c,i}- RFE_{min\ c,i})$ | Predictor variable: Normalized RFE to the range [0,1] for each extent and for each time period |
| SPI | $SPI_{c,i}=(RFE_{c,i}-RFE_{mean\ c,i})/ RFE_{stdev\ c,i}$ | Predictor variable: Standardised RFE for each extent and for each time period. |

In the variable transformation step, we generated the 1-3-month lags of each of the variables for each county. The non-lagged variables are then dropped from the study. The resultant dataset is as described in Table 3. Random sampling is used to partition the data into training and validation datasets. This follows, on the 70:30 rule for training and validation data sets, respectively.

**Table 3:** Variables used for modelling

| Index | Variable description | 1-Month Lag | 2-Month Lag | 3-Month Lag |
|---|---|---|---|---|
| *NDVI_Dekad* | *NDVI for last dekad of month* | ☒ | ☒ | ☒ |
| *VCI_Dekad* | *VCI for the last dekad of month* | ☒ | ☒ | ☒ |
| *VCI1M* | *VCI aggregated over 1 month* | ☒ | ☒ | ☒ |
| *VCI3M* | *VCI aggregated over the last 3 months* | ☒ | ☒ | ☒ |
| *RFE1M* | *Rainfall Estimate aggregated over 1 month* | ☒ | ☒ | ☒ |
| *RFE3M* | *Rainfall Estimate aggregated over the last 3 months* | ☒ | ☒ | ☒ |
| *SPI1M* | *Standardised Precipitation Index aggregated over 1 month* | ☒ | ☒ | ☒ |
| *SPI3M* | *Standardised Precipitation Index aggregated over the last 3 months* | ☒ | ☒ | ☒ |
| *RCI1M* | *Rainfall Condition Index aggregated over 1 month* | ☒ | ☒ | ☒ |
| *RCI3M* | *Rainfall Condition Index aggregated over the last 3 months* | ☒ | ☒ | ☒ |
| *Month* | *Denotes the month of above variables. Used to model seasonality* | ☒ | ☐ | ☐ |

## 2.3 Methods

The study uses multiple indices and combines multiple methods in the prediction of drought. The prediction of drought is, for operational purposes, formulated as the prediction of future VCI values using the above presented predictor variables (Table 3). We focus here on predictions 1-month ahead, while 3-month ahead predictions were also tested. Two approaches are combined (Figure 3): a statistical approach - Generalized Linear Models (GAM) and Artificial Neural Networks (ANN). GAM models are used to arrive at the set of variables that offer the best predictions. These set of variables are then used to build ANN models. GA models are thus used as a model variables selection method to the ANN modelling. GA models are reviewed for example in [98] and a good description of ANNs is provided in YYY [99].

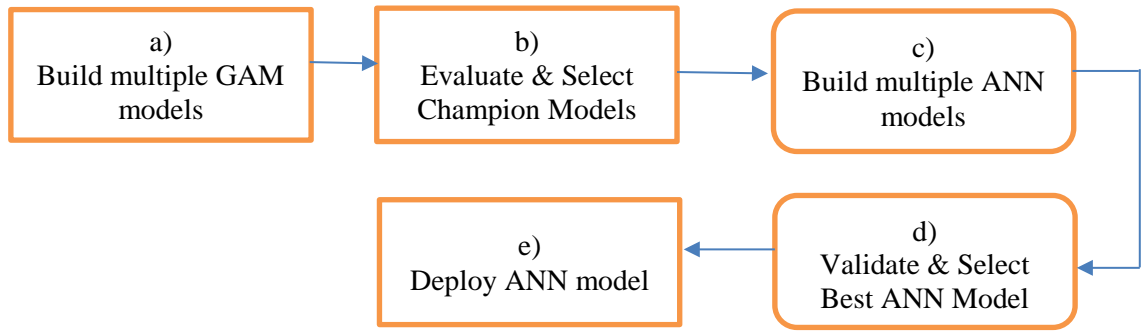

**Figure 3:** Schema of the modelling process. In the sub-processes a) and b), GAM models are used to arrive at the set of variables that offer the best predictions. These set of variables that are then used to build ANN models. GA models are thus used, essentially as a model variable selection method to the subsequent ANN modelling sub-processes.

The model space reduction process is used to reduce on the model selection complexity. The assumptions and rules used for model space reduction are presented in Figure 4. The final model space for evaluation is therefore made up of 102 models for the GAM modelling process.

| $\sum_{r=1}^{31} \frac{n!}{(n-r)!\,r!}$ | a) | $\sum_{r=1}^{2} \frac{n!}{(n-r)!\,r!}$ | b) | 102 |
|---|---|---|---|---|
| U=2,147,483,647 | | 496 | | |

**Figure 4:** Results of the model space reduction process with an initial model space of almost 2.15 billion. Enforcement of an assumption of sufficiency of two variables plus a third for seasonality a) reduces the model space to 496 models. A further reduction is achieved by ensuring no two variables of the same type (precipitation and vegetation) are used in the same model and also ensuring only one level of lag variables are used in the same model. This reduces the space of models to 102.

### 2.3.1 Generalized Additive Models (GAM)

GAM models were selected because they do not assume linearity between the predictor and the response variables [REF]. In addition, GAM are almost free form since they do not require the ascertainment of the functional form of relationship to be modelled beforehand. In the case that the relationships are best approximated by linear, quadratic or cubic functions, GAM results simplify to these as is appropriate. These coupled with the fact that we still have the desirable features of GLiMs and GLMM make GAM models a viable tool for weather-based data modelling.

GAM models are expressed as shown in Formula 1.

$$Y = a + f_1(x_1) + f_2(x_2) + \cdots + f_n(x_n) + \varepsilon$$

where *a* is an intercept and *f* are smooth functions; *Y* is the response function and $x_1$ to $x_n$ are the n predictor variables.

Smoothing functions are either local linear regression (loess) or splines. In practical application, caution is advised since smoothing generally leads to model overfitting.

The space of models for the study, as given by Equation 1, is around 2.15 Billion. This space would be impractical to navigate in the search for the best predictor model.

$$\sum_{r=1}^{31} \frac{n!}{(n-r)!\,r!} \approx 2.15B$$

To minimise the space complexity, we make some *a priori* assumptions to avoid the futility of bias-free learning and also follow Ocam's razor Mitchell (1997) [27]. First, we assume that a maximum of two variables in the GAM models will give us reasonably simple models while still remaining predictive. Second, we assume that including multiple variables of the same category (vegetation or precipitation) is an unnecessary increase of the complexity of the model space at marginal possible increased performance. Together with these two assumptions, we further use a rule of thumb to not use different lag times of the same variable in a single model. To capture seasonality, we further include an additional variable for the month of the year of the instances as a sine wave. Seasonality is expected to exist in precipitation and vegetation cover data.

With the model space reduced to 102 (Figure 4), we brute-forced the process of training and evaluating the models in an automated process. Multiple model evaluation metrices were used and the results logged for both model training and model evaluation. Results are reported separately for training and validation data.

*2.3.2 Artificial Neural Networks*

Artificial Neural Networks (ANNs) are a Machine Learning (ML) approach that mimic the interconnectedness of the brain in the modelling process Mitchell (1997) [27]. ANNs are capable of learning discrete, real and vector valued functions and vastly remain robust to errors in training data.

ANNs have several characteristics making them suitable for the purpose of predictive modelling: (1) instances can be represented by many attribute value pairs, (2) the target function is either discrete, real or vector valued, (3) training examples may contain errors, (4) non-linear relations can be modeled, and (5) execution (after training) is very quick, …(4) long training times are acceptable while faster evaluation is required since the process will be run on a monthly basis.

To overcome overfitting which is the most common limitation of ANNs, we chose models that are judged to perform better in the evaluation datasets as compared to the training datasets using $R^2$ as the measure of model performance. Our working definition of overfitting is presented in Formula 2 together with the General Additive Model (GAM) results.

The ANNs were built using the back-propagation algorithm and for the limitation of complexity, the modelling process was set to have a formation of 2-5-3-1 and thus had two hidden layers that are able to learn any arbitrary function. The formation was realised from both a rule of thumb and an experimentation process. The rule of thumb is based on Formula 2 from Huang (2003) [28]. The total number of hidden nodes was thus set at eight.

$$Number\ of\ nodes = \begin{cases} Sqrt(N*(m+2)) + 2*Sqrt(N/(m+2)), & hidden\ layer\ one \\ m*Sqrt\ (N/(m+2)), & hidden\ layer\ two \end{cases} \quad \text{--Formula (2)}$$

The maximum step was set to 1e+06 and this represented the maximum steps for the training of the neural network at whose attainment the network's training process is stopped. The maximum step size was a failsafe condition for the ANNs, should the pre-selected set of hidden layers not lead to convergence majorly due to partitions in the training and validation datasets.

The process for the execution of the artificial neural networks modelling is as presented in Figure 5 and was built on normalised variables. Variable normalization was done prior to model training and let to values in the [0,1] range. The values were centred at the minimum value for each variable, then scaled between the minimum and maximum values.

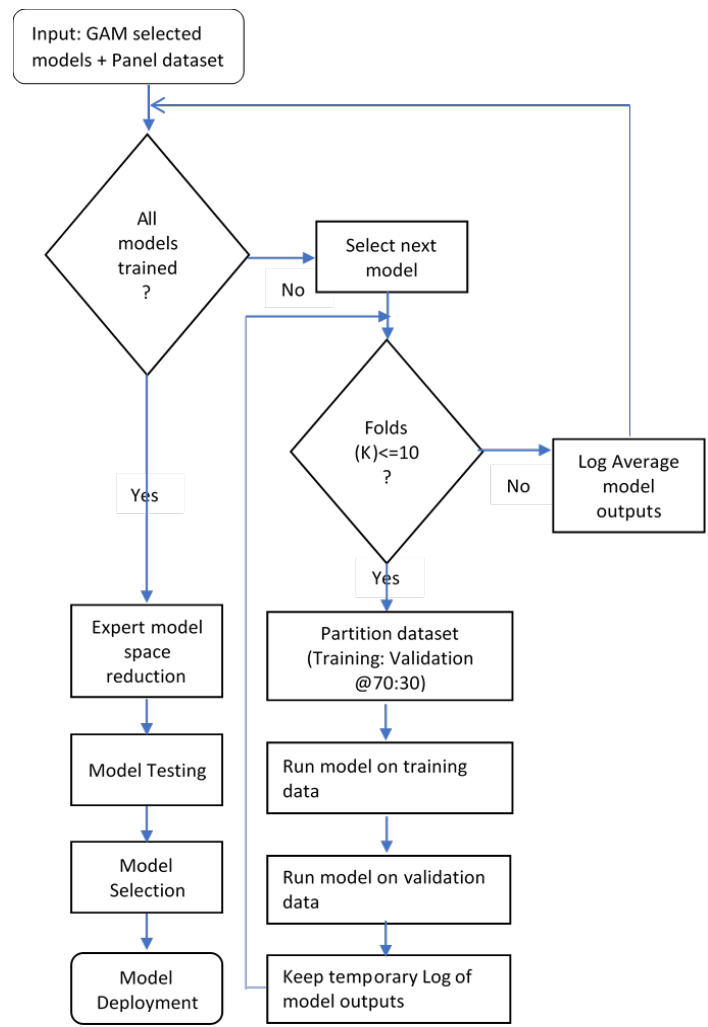

**Figure 5:** Outline of the ANN modelling process. The process sequentially inputs selected GAM models and the panel dataset followed by the iteration of the performance of the models against the data. The data is randomly partitioned in the ratio 70:30 for training and validation, respectively, for each and every iteration of the k times a model is run against the data. The k-fold iteration was chosen to minimize impacts of the random initialisation of the network weights.

## 3. Results and Discussion

### 3.1 GAM Model results

For all the models run, and for both GAM and ANN, the validation metrics used are mean absolute error (mae), mean squared error (mse), root mean squared error (rmse), mean percentage error (mape), nmean squared error (nmase), name and r-squared.

Each GAM model had the formula modified to an extra variable that is the smoothed sine function of the month number that is used to capture seasonality of the vegetation data. The smoothing uses cyclic cubic regression splines (cc) that has start and end points and thus appropriate for modelling seasonality.

A plot of the performance of the 102 models in the GAM process, grouped by $R^2$ is presented in Figure 6. The models are noted to post $R^2$ between 0.09 and 0.86 in model training and model validation. The performances of the models in training and validation datasets (blue and orange bars respectively) indicate relative stability in model numbers across the models.

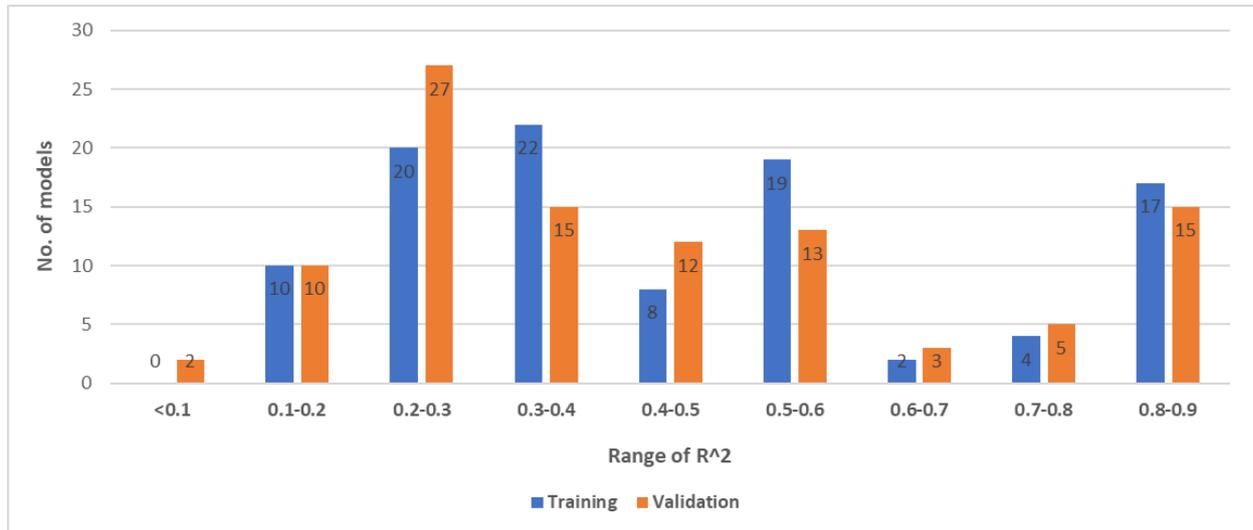

**Figure 6:** Model performance by range of $R^2$ in the GAM process.

The performance of the models by the lag-time of the variables (between 1- and 3-month lags) is provided in Figure 7. As expected, the analysis of the GAM process by lag time indicates that the 1-month lag of the predictors perform better in predicting VCI3M as used to define drought (in green). While a lag time of 2 month (in blue) still has some strong predictive power ($R^2 >0.5$), even longer lags fail to produce good correlations (in orange).

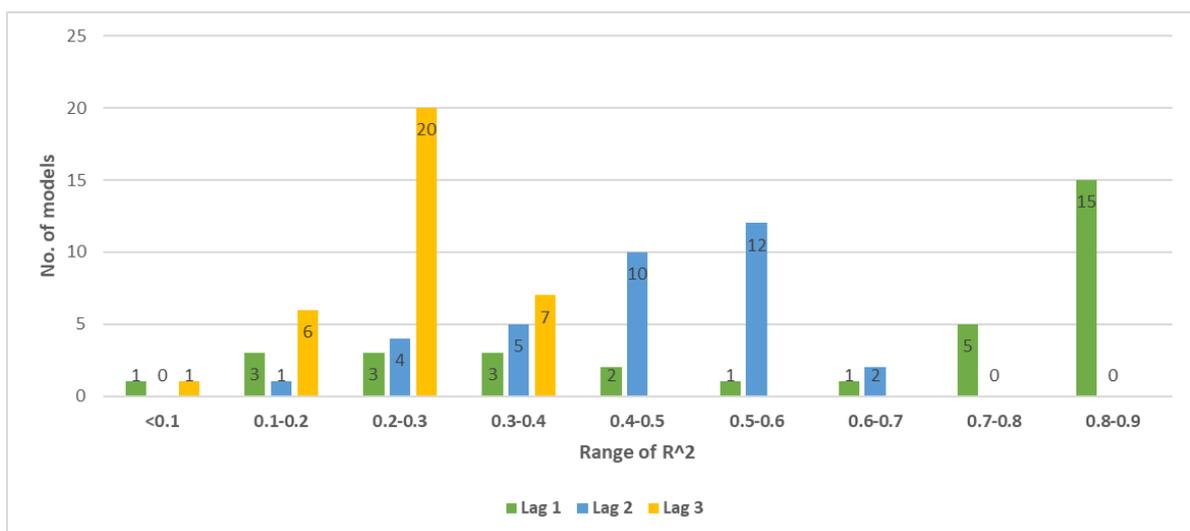

**Figure 7:** Lag-time based performance of the GAM model selection space reduction process

It is deducible that the GAM process models with $R^2 \geq 0.7$ (Table 4) are all 1-month lag variables. In fact, the first 2-month lag variable first appears at a model ranked at position 22 with an $R^2$ of 0.61 while the first 3-month lag variable is in a model ranked at position 52 with an $R^2$ of 0.33. The poor performance of higher lags of these variables is expected since longer lags are less contributing to current vegetation status – and unexpected climate variations may occur between time of forecasting and the forecasted event.

**Table 4:** Models with $R^2$-squared $\geq 0.7$. Also indicated are models indicating on overfitting. Even though the model at position 21 is indicated to have an R2 of 0.69, rounded off for comparison with 0.7, it is chosen as meeting the criteria for selection by the GAM process.

| No | Model | $R^2$ Training | $R^2$ Validation | Overfit Index | Overfit | Lag Time |
|---|---|---|---|---|---|---|
| 1 | VCIDekad_lag1+SPI1M_lag1 | 0.86 | 0.85 | 0.01 | No | 1 |
| 2 | VCIDekad_lag1+SPI3M_lag1 | 0.86 | 0.85 | 0.01 | No | 1 |
| 3 | VCIDekad_lag1+RFE1M_lag1 | 0.85 | 0.85 | 0.01 | No | 1 |
| 4 | VCI1M_lag1+SPI3M_lag1 | 0.85 | 0.84 | 0.01 | No | 1 |
| 5 | VCI1M_lag1+SPI1M_lag1 | 0.85 | 0.84 | 0.01 | No | 1 |
| 6 | VCI1M_lag1+RFE1M_lag1 | 0.85 | 0.84 | 0.01 | No | 1 |
| 7 | VCIDekad_lag1+RCI1M_lag1 | 0.85 | 0.84 | 0.01 | No | 1 |
| 8 | VCI1M_lag1+RCI1M_lag1 | 0.84 | 0.83 | 0.01 | No | 1 |
| 9 | VCIDekad_lag1+RCI3M_lag1 | 0.84 | 0.83 | 0.01 | No | 1 |
| 10 | VCIDekad_lag1+RFE3M_lag1 | 0.84 | 0.83 | 0.01 | No | 1 |
| 11 | VCI1M_lag1+RCI3M_lag1 | 0.84 | 0.83 | 0.01 | No | 1 |
| 12 | VCI1M_lag1+RFE3M_lag1 | 0.83 | 0.83 | 0.01 | No | 1 |
| 13 | VCI3M_lag1+SPI3M_lag1 | 0.82 | 0.82 | 0.01 | No | 1 |
| 14 | VCIDekad_lag1 | 0.81 | 0.80 | 0.01 | No | 1 |
| 15 | VCI3M_lag1+RCI3M_lag1 | 0.81 | 0.80 | 0.01 | No | 1 |
| 16 | VCI1M_lag1 | 0.81 | 0.80 | 0.01 | No | 1 |
| 17 | VCI3M_lag1+SPI1M_lag1 | 0.81 | 0.79 | 0.01 | No | 1 |
| 18 | VCI3M_lag1+RCI1M_lag1 | 0.78 | 0.77 | 0.01 | No | 1 |
| 19 | VCI3M_lag1+RFE3M_lag1 | 0.78 | 0.77 | 0.01 | No | 1 |
| 20 | VCI3M_lag1+RFE1M_lag1 | 0.78 | 0.76 | 0.01 | No | 1 |
| 21 | VCI3M_lag1 | 0.72 | 0.69 | 0.02 | No | 1 |

With the definition of overfitting based on Equation 3, it is shown that none of the 21 GAM models with R-squared greater than 70% are judged to have suffered over-fitting. All the 21 models are thus noted to have acceptable deterioration in performance in validation.

$$Overfit\ model = \begin{cases} Yes, & diff(RsquaredT, RsquaredV) \geq 0.03 \\ No, & otherwise \end{cases} \quad --Formula\ (3)$$

where $RsquaredT = R-squared\ in\ training\ set$ and $RsquaredV = R-squared\ in\ validation\ set$

A final analysis was done on the effect of smoothing all the variables in the GAM modelling process. This resulted in highly overfit models with 13 of the above top 21 models (62%) overfit with up to an overfit index of 0.07 (7%) (not shown). This is a known problem of smoothing in GAM models. Despite the tendency to overfit the models, it is worth noting that the same set of models are ranked highly by the GAM modelling technique.

The alternative measures of performance (MAE, MSE, RMSE, MAPE, NMSE and NAME) are noted to be consistent with R squared since they all have a non-monotonic and non-linear relationships. An increase in $R^2$ translates to a change but in reverse direction of the other measures of model performance.

Since the aim of the study was to use GAM modelling process as a basis for model space reduction, the above 21 models were selected for the ANN process.

## 3.2 ANN Model results

The intent of the study is to use ANN's as the case study technique of choice. Following on the model space search approach, we produced all the 21 models using the ANN process through a bagging and brute force approach in the search for the champion model. For uniformity, overfitting is defined for Artificial Neural Networks (ANN) as in GAM models.

### 3.2.1 ANN Performance in Training and Validation

Using the model overfit index defined earlier on the model statistics for the Neural Networks (Table 5), a few facts emerge. The ANN models are generally not overfit with only one model (No. 19) overfit. This implies a non-overfit rate of 95%.

**Table 5:** ANN model performances in training and validation datasets. In gray, the only overfitting model

| No | Model | Training ($R^2$) | | | Validation ($R^2$) | | | Overfit Index | Overfit |
|---|---|---|---|---|---|---|---|---|---|
| | | Min | Max | Mean | Min | Max | Mean | | |
| 1 | VCIDekad_lag1+RFE1M_lag1 | 0.83 | 0.86 | 0.84 | 0.78 | 0.86 | 0.83 | 0.010 | No |
| 2 | VCI1M_lag1+RFE1M_lag1 | 0.82 | 0.85 | 0.84 | 0.78 | 0.85 | 0.83 | 0.010 | No |
| 3 | VCIDekad_lag1+SPI1M_lag1 | 0.82 | 0.85 | 0.84 | 0.79 | 0.87 | 0.82 | 0.017 | No |
| 4 | VCIDekad_lag1+SPI3M_lag1 | 0.82 | 0.86 | 0.84 | 0.78 | 0.88 | 0.82 | 0.021 | No |
| 5 | VCIDekad_lag1+RCI3M_lag1 | 0.82 | 0.86 | 0.84 | 0.79 | 0.87 | 0.82 | 0.020 | No |
| 6 | VCI1M_lag1+SPI3M_lag1 | 0.81 | 0.85 | 0.84 | 0.78 | 0.87 | 0.82 | 0.018 | No |
| 7 | VCI1M_lag1+RCI3M_lag1 | 0.82 | 0.85 | 0.84 | 0.79 | 0.86 | 0.82 | 0.021 | No |
| 8 | VCI1M_lag1+SPI1M_lag1 | 0.82 | 0.85 | 0.84 | 0.77 | 0.86 | 0.82 | 0.019 | No |
| 9 | VCIDekad_lag1+RCI1M_lag1 | 0.81 | 0.84 | 0.82 | 0.76 | 0.85 | 0.81 | 0.017 | No |
| 10 | VCI1M_lag1+RCI1M_lag1 | 0.80 | 0.84 | 0.82 | 0.75 | 0.84 | 0.80 | 0.016 | No |
| 11 | VCIDekad_lag1+RFE3M_lag1 | 0.79 | 0.84 | 0.82 | 0.75 | 0.83 | 0.80 | 0.016 | No |
| 12 | VCI1M_lag1+RFE3M_lag1 | 0.79 | 0.84 | 0.81 | 0.74 | 0.83 | 0.79 | 0.017 | No |
| 13 | VCIDekad_lag1 | 0.77 | 0.82 | 0.79 | 0.72 | 0.82 | 0.78 | 0.014 | No |
| 14 | VCI1M_lag1 | 0.76 | 0.81 | 0.78 | 0.72 | 0.81 | 0.77 | 0.015 | No |
| 15 | VCI3M_lag1+SPI3M_lag1 | 0.76 | 0.81 | 0.79 | 0.73 | 0.84 | 0.77 | 0.027 | No |
| 16 | VCI3M_lag1+RFE1M_lag1 | 0.76 | 0.79 | 0.77 | 0.72 | 0.80 | 0.77 | 0.006 | No |
| 17 | VCI3M_lag1+RCI3M_lag1 | 0.76 | 0.81 | 0.79 | 0.72 | 0.83 | 0.76 | 0.027 | No |
| 18 | VCI3M_lag1+RCI1M_lag1 | 0.74 | 0.79 | 0.77 | 0.71 | 0.80 | 0.75 | 0.023 | No |
| 19 | VCI3M_lag1+SPI1M_lag1 | 0.73 | 0.80 | 0.78 | 0.70 | 0.82 | 0.74 | 0.036 | Yes |
| 20 | VCI3M_lag1+RFE3M_lag1 | 0.71 | 0.77 | 0.74 | 0.65 | 0.76 | 0.72 | 0.019 | No |
| 21 | VCI3M_lag1 | 0.64 | 0.71 | 0.68 | 0.60 | 0.73 | 0.66 | 0.023 | No |

Since the methodology used runs the same ANN model across 10 different partitions of the data sets, a review of model results indicates that almost all the models post an $R^2$ of at least 0.7 in all the partitions of the training data except for two models (No 20 and 21 in Table 5).

The champion model from the ANN process is different from that of the GAM process. In fact, the ANN champion ($R^2$=0.83) was ranked the third best model in the GAM modelling process ($R^2$=0.85). Figure 8 illustrates the performance of the ANN models as compared to the GAM models.

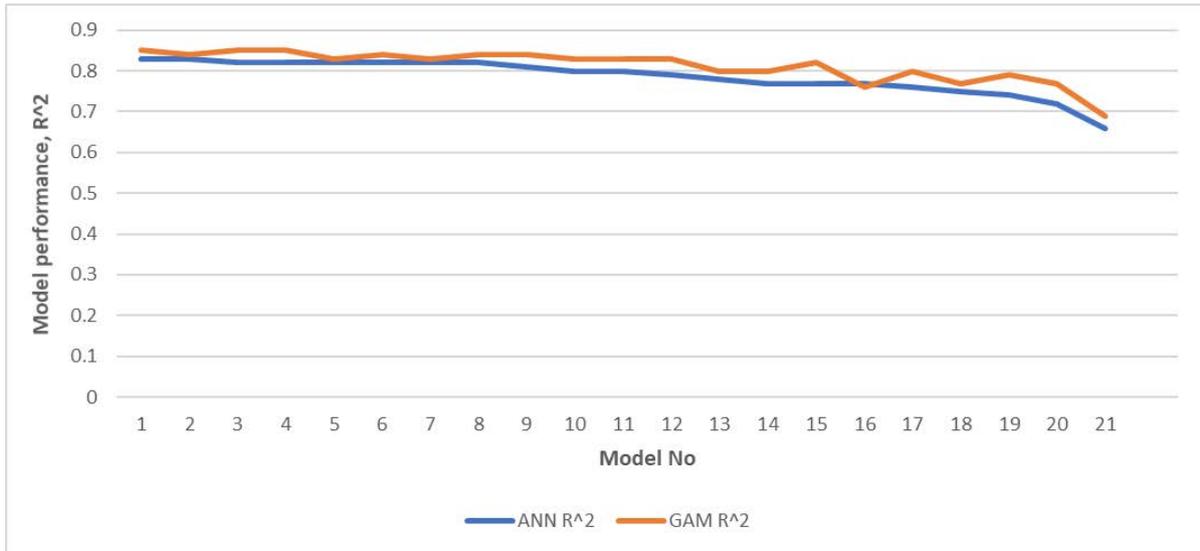

**Figure 8:** Performance of the ANN models in validation as compared to similar GAM models

In general, as indicated in Figure 8, the GAM models outperform ANN models except for model 16 for which the ANN slightly out-performs GAM by an $R^2$ of 0.01. This is an important property since the GAM process is proved to be more optimistic in performance as compared to ANN and so fewer deserving models would be excluded from the ANN process.

In the training and validation, the champion model that has its best subset performance (Max $R^2$=0.86) presented in Figure 9, has a positive upward trend between the actual and the predicted values for the vegetation condition index.

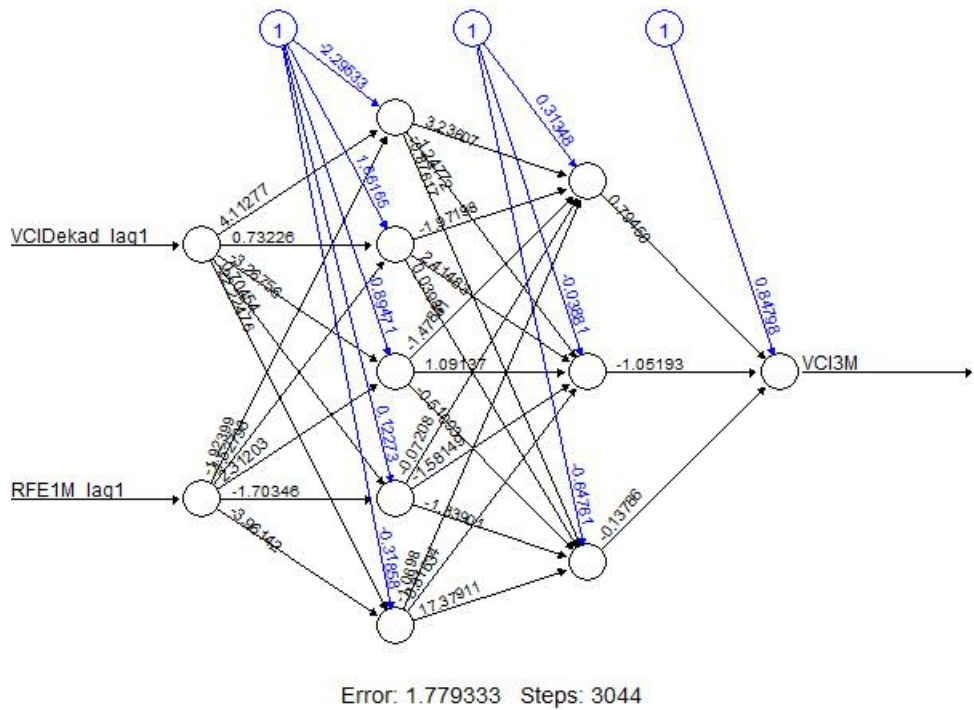

**Figure 9:** The Champion model with the variables VCIDekad_lag1 and RFE1M_lag1. The plot is from the 4[th] partition that recorded the best performance. modelling method with the variables . Blue lines indicate bias values, and black lines internal weights

*3.2.2 ANN Champion model's performance in test dataset*

The out-of-sample test dataset has 96 data points across a 2-year period. The out-of-sample data was neither used in the training nor the validation processes of the ANN and even of the GAM process. It represents the model's performance in the real world.

*Performance of ANN in Regression*

The ANN prediction was first formulated as a regression. The performance of the ANN champion in regression is indicated in the plot of the actuals versus the predicted real values as shown in Figure 10.

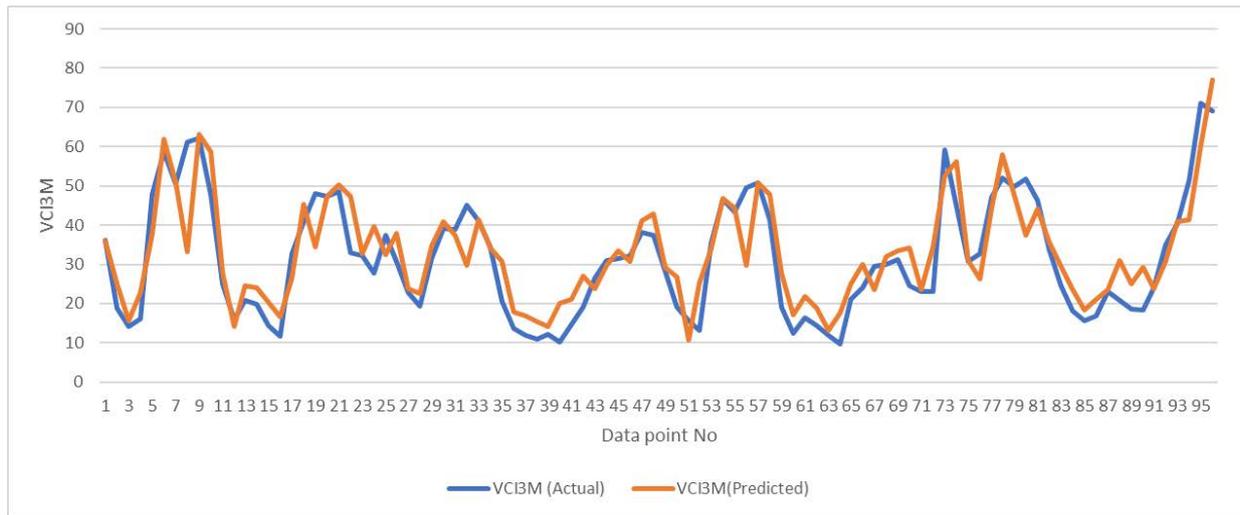

**Figure 10:** Plot of the actual versus champion model's predicted values in test data for Turkana county. Predictions were done 1 month ahead (results for the other three counties are shown in Annex)

The plot of the actual versus the predicted values represents quite a good agreement. In the test data, the champion models posted an $R^2$ of 0.78 and RMSE of 7.03 on the actual data values. The above performance over the 96 data points for testing is an acceptable performance in the prediction of drought events.

*Performance of ANN in classification*

Operational drought monitoring involves the definition of thresholds on indices used for drought monitoring so as to realise a phase approach to drought monitoring. We use the approach documented in Klisch and Atzberger (2016) and Meroni et al. (2019) in Table 6 to monitor drought in five phases.

**Table 6:** Phase classification of drought based on vegetation deficit following values proposed by Klisch and Atzberger (2016) and Meroni et al. (2019)

| VCI3M Lower limit | VCI3M Upper limit | Description of Class | Drought class |
|---|---|---|---|
| 0 | 10 | Extreme vegetation deficit | 1 |
| 10 | 20 | Severe vegetation deficit | 2 |
| 20 | 35 | Moderate vegetation deficit | 3 |
| 35 | 50 | Normal vegetation conditions | 4 |
| 50 | 100 | Above normal vegetation conditions | 5 |

| Month No | 1 | 2 | 3 | 4 | 5 | 6 | 7 | 8 | 9 | 10 | 11 | 12 | 13 | 14 | 15 | 16 | 17 | 18 | 19 | 20 | 21 | 22 | 23 | 24 |
|---|---|---|---|---|---|---|---|---|---|---|---|---|---|---|---|---|---|---|---|---|---|---|---|---|
| Actual | | | | | | | | | | | | | | | | | | | | | | | | |
| Predicted | | | | | | | | | | | | | | | | | | | | | | | | |
| Difference | | | | | | | | | | | | | | | | | | | | | | | | |

**Figure 11:** Performance of the classifier for the case of Turkana county showing months of difference in grey and those of agreement in blue.

The champion model had an overall accuracy of 66% rising to 71% for Turkana county as indicated in matrix provided in Figure 11.

When formulated as a multi-class classification problem, the one-versus all plot of the ROC for the 5 classes in Figure 10, provides a reasonable trade-off between sensitivity and specificity at an overall area under the multi-class ROC (AUROC) of 89.99% as defined by Hand and Till (2001).

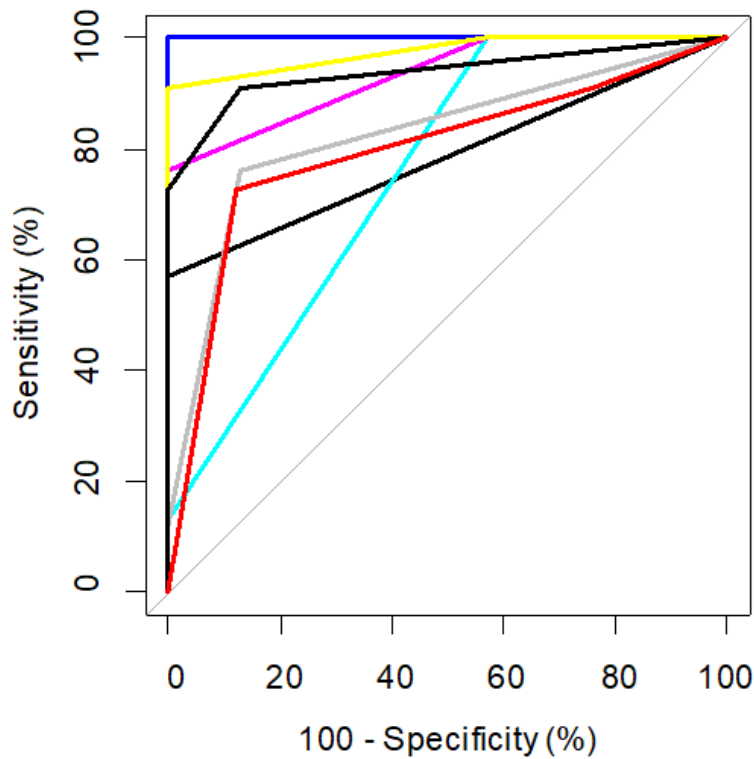

**Figure 12:** Multi-class ROC plot of the champion model as a drought phase classifier.

The area under the ROC (AUROC) indicates quite a good trade-off between sensitivity and specificity and is ranked within the good performance category as it is way above the 50% that represents a worthless test.

### 3.2.3 Validation of the key assumption of the study

To validate the key assumption on the appropriateness of the GAM modelling technique in the reduction of the model space we run the extra 81 models through the ANN process. The best performer from the set of non-selected models had an $R^2$ of 0.50. A summary of the performance of the non-selected models in the test dataset is provided in Figure 13.

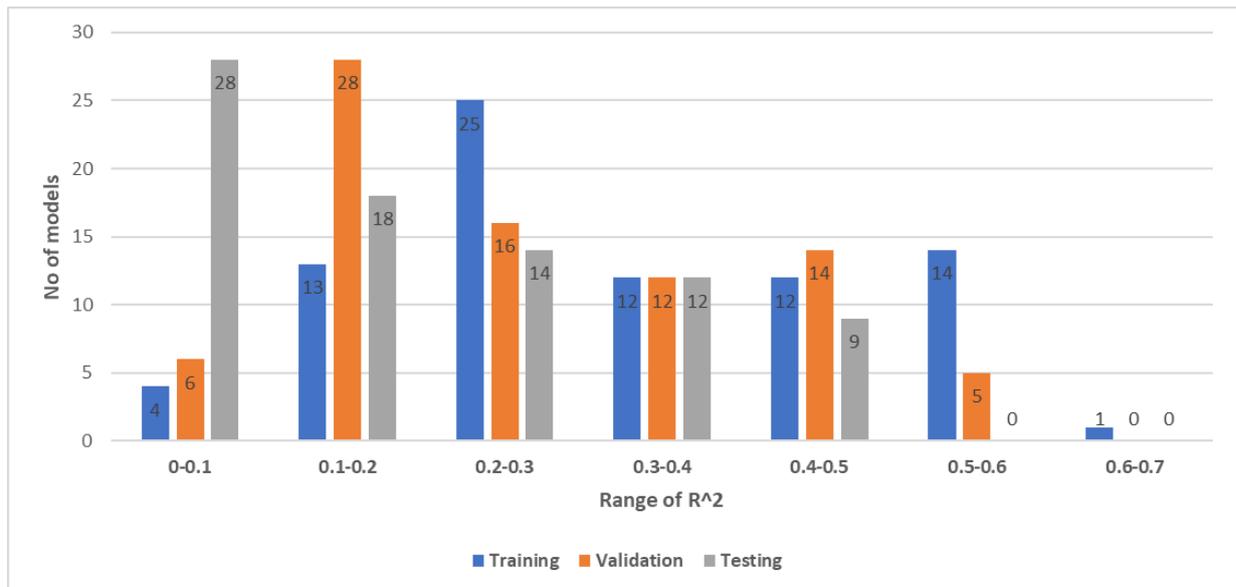

**Figure 13:** Distribution of non-selected models ANN performance on training, validation and testing. No model is noted to post an $R^2$ of at least 0.5 in testing.

The results validate the assumption of the utility of GAM in modelling non-linearity as well as in the possibility of their use in model space reduction before the use of computationally intensive algorithms like artificial neural networks. The models non-selected by ANN are expected not to perform any better using ANNs.

## 4. Conclusion

In this paper, multiple variables that are transformations from primary indicators of rainfall and vegetation cover are used to predict future vegetation condition index (VCI) as a proxy to drought conditions. The methodology uses two techniques in a setup where the General Additive Model (GAM) statistical approach is first run against several possible model configurations. The GAM method is then used to reduce the model space and by extension the set of viable variables. After variable selection and with the model space reduced, a brute force approach is then employed using the Artificial Neural Networks (ANN) approach. The model space reduction is beneficial to the building of neural networks that are known to generally have slower training times as compared to other approaches. The automation of the model training and model validation processes and the measure of performance with a view to quantifying and avoiding overfitting make for a scalable approach. The poor performance of variables with longer times lags in the prediction of drought events is established and the moderate performance of reformulation and evaluation of regressors in-terms of their equivalent classification outputs. The approach builds multiple models and can thus be used in a model ensembling approach to prediction of future drought conditions.